\documentclass[12pt]{amsart}

\usepackage[margin=2.5cm]{geometry}
\usepackage{tikz}
\tikzset{node distance=2cm, auto}
\usetikzlibrary{matrix,arrows,decorations.pathmorphing}

\usepackage{graphicx}
\usepackage{amsmath,amssymb}
\newcommand\numberthis{\addtocounter{equation}{1}\tag{\theequation}}
\usepackage{amscd}
\usepackage{verbatim}
\usepackage{enumerate}

\newtheorem{theorem}{Theorem}

\newtheorem{proposition}{Proposition}

\newtheorem{remark}{Remark}
\newtheorem{conjecture}{Conjecture}

\newcommand{\g}{\mathfrak{g}}

\newcommand{\RR}{\mathbb{R}}

\newcommand{\ZZ}{\mathbb{Z}}

\newcommand{\pa}{\partial}

\newcommand{\sign}{\operatorname{sign}}

\newcommand{\be}{\begin{equation}}
\newcommand{\ee}{\end{equation}}

\newcommand{\cL}{\mathcal{L}}

\newcommand{\cM}{\mathcal{M}}

\newcommand{\cO}{\mathcal O}

\newenvironment{dedication}
        {\vspace{6ex}\begin{quotation}\begin{center}\begin{em}}
        {\par\end{em}\end{center}\end{quotation}}

\title{Semiclassical geometry of integrable systems.}
\author{Nicolai Reshetikhin}
\address{N.R.: Department of Mathematics, University of California, Berkeley,
CA 94720, USA \& Physics Department, St. Petersburg University, Russia \&KdV Institute for Mathematics, University of Amsterdam,
Science Park 904, 1098 XH Amsterdam, The Netherlands.}
\email{reshetik@math.berkeley.edu}
\begin{document}
\begin{dedication}
\hspace{4cm}
\vspace*{3cm}{Dedicated to the memory of P.P.Kulish.}
\end{dedication}
\begin{abstract}
The main result of this paper is a  formula for the scalar product of semiclassical eigenvectors
of two integrable systems on the same symplectic manifold. An important application of this formula
is the Ponzano-Regge type of asymptotic of Racah-Wigner coefficients.
\end{abstract}
\maketitle

\section*{Introduction}

One of the main motivations for this paper is the study of semiclassical asymptotics of $6j$-symbols
and their quantum analogs. Recall that $6j$ symbols, which are also known as Racah-Wigner coefficients,
first appeared in the theory of angular momentum. They describe the transition matrix between two
natural bases in the decomposition of the tensor product of three irreducible
representations of $SU(2)$ into irreducibles. More generally, they describe the associators
in braided monoidal categories which naturally appear in representation theory
of simple Lie algebras or corresponding quantum groups.

The semiclassical asymptotic in representation theory is the limit when all
components of highest weights go to infinity at the same rate. In this limit
many features of representation theory can be expressed in terms of geometry of coadjoint orbits.
For $SU(2)$ the semiclassical asymptotic of $6j$ symbols was computed
by Ponzano and Regge in \cite{PR}.
A geometric interpretation of Ponzano-Regge asymptotic, involving symplectic
geometry was done by Roberts in \cite{Ro}, where he observed that this is
essentially the computation of the semiclassical asymptotic of the scalar product
of eigenfunctions of Hamiltonians of two integrable systems.
Similar observation was done by Taylor and Woodward in \cite{TW1} \cite{TW2} where the authors
computed the semiclassical
asymptotic for q-6j symbols \cite{KR}. For references to earlier works see \cite{PR}\cite{Ro}.
See also \cite{Ch} where the approach of \cite{Ro} was extended to the tensor product of
$N$ representations.

In this paper we give the general formula (\ref{wkb2systq}) and (\ref{general})  for leading terms of the semiclassical asymptotic of
the scalar product of eigenfunctions of two integrable systems on the
same phase space. It is closely related to the quantization
of Lagrangian submanifolds, see for example \cite{BW}\cite{GS}\cite{GS2}.
In the setting of the geometric quantization eigenfunctions should be understood
as half-densities, so it is better to say that we study scalar products of eigen-half-densities.
For the spectral analysis of quantum integrable systems see \cite{PS} and references therein.

In particular the formula (\ref{general}) for the scalar product gives the semiclassical asymptotic for
$6j$ symbols (and $q$-6j symbols) for all simple Lie algebras in multiplicity free cases. We will
not focus on this particular application here, but will address it in a separate
publication.

Another natural place where similar formula appears is the semiclassical asymptotic of the
propagator in quantum mechanics when the initial quantum space of states and
the target space of states are described in terms
of geometric quantization when they correspond to different, transversal real polarizations of
the phase space.  This is discussed in the Conclusion.

The plan of the paper is as follows. The first section is focused on the one-dimensional case. In the second section
the formula is derived for two integrable systems on a cotangent bundle.
In the Conclusion the semiclassical formula is stated for general symplectic manifolds, the relation to
$6j$ symbols is discussed and the relation to topological quantum mechanics is outlined.

The author is grateful to J. E. Andersen, A.Cattaneo, S. Dyatlov, V. Fock, P. Mnev, L. Polterovich, S. Shakirov, B. Tsygan, and M. Zworski for stimulating discussions.
This work was supported by the NSF grant DMS-1601947.

\section{The one dimensional example}
\subsection{Hamiltonians which are quadratic in momentum} Here we recall some basic facts about the semiclassical, WKB\footnote{Recall that WKB stands for Wentzel--Kramers--Brillouin.} asymptotic
of eigenfunctions for the one dimensional Schrodinger operator.

Let $H(p,q)=\frac{p^2}{2}+V(q)$ be the Hamiltonian of a classical system describing a one dimensional particle with mass $1$
in an external potential $V(q)$ which
we assume to be a smooth function. This Hamiltonian is a function on the phase space $\RR^2=T^*\RR$ which is equipped with the standard symplectic
form $\omega=dp\wedge dq$, where $p$ is the coordinate along the cotangent fibers and $q$ is a coordinate on the configuration space $\RR$.
Generic level curves of $H$ are Lagrangian submanifolds in $\RR^2$. For simplicity we will focus on the
case when level curves are connected and compact\footnote{In case when level curves are not connected we will have to consider tunneling between
different components. When they are noncompact there is no Bohr-Sommerfeld quantization condition.}.

\begin{figure}[htb]
\includegraphics[height=6cm,width=8cm]{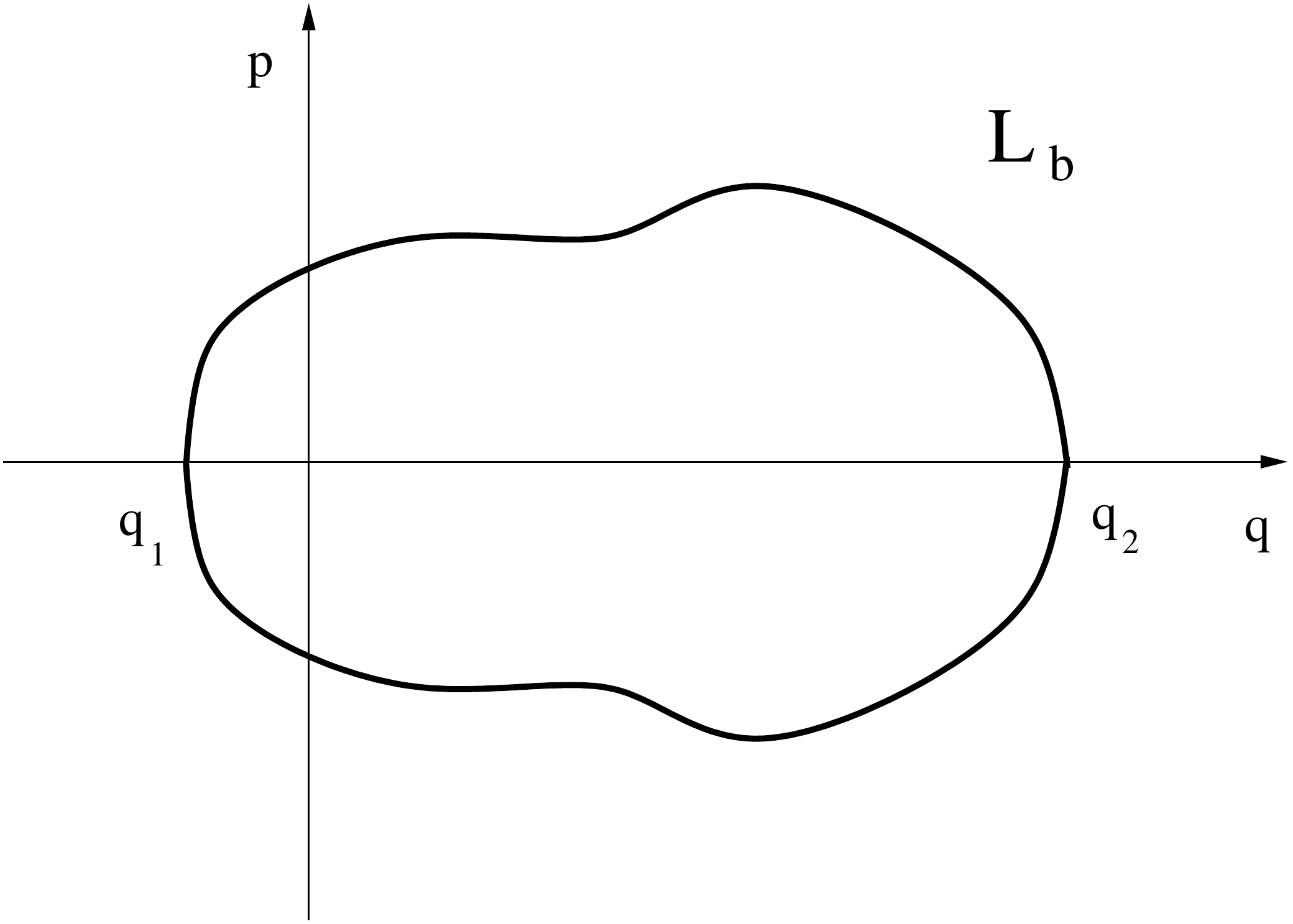}
\caption{Level curve of a quadratic Hamiltonian with turning points.}
\label{F1}
\end{figure}

With these assumptions the level curve  $H(p,q)=b$
\begin{equation}\label{E-surf}
\cL_b=\{(p,q)|\frac{p^2}{2}+V(q)=b\}
\end{equation}
is a double cover of the segment $q_1<q<q_2$
where $q_i$ are turning points $V(q_i)=b$ (see Fig. \ref{F1}).

We use $b$ for the energy because we consider it as a point of the Lagrangian fibration
$H: \RR^2\to \RR$ given by level curves of $H(p,q)$.

Denote by $p_\pm(q,b)$ two branches of this  level curve
$p_\pm(q,b)=\pm p(q,b)$ where $p(q,b)=\sqrt{2(b-V(q))}$ and by $S_\pm(q,b)$ corresponding branches of the Hamilton-Jacobi action function:
\[
\frac{\pa S_\pm(q,b)}{\pa q}=\pm p(q,b)
\]
Here $q$ are two points on $\cL_b$ over $q\in \RR$.

Quantized of a classical Hamiltonian $H(p,q)$  is a differential operator, with $H(p,q)$ being its principal symbol.
The Schroediner operator
\[
\widehat{H}=-\frac{h^2}{2} \frac{d^2}{dq^2}+V(q)
\]
is the physically important quantization\footnote{Here $h$ is the Plank constant in appropriate units.}.

The semiclassical, i.e. $h\to 0$, asymptotic of eigenfunctions of the quantized Hamiltonian
\begin{equation}\label{1d-schr}
\widehat{H}\psi_b(q)=\widehat{b}\psi_b(q)
\end{equation}
is well known as the  WKB asymptotic:
\begin{equation}\label{wkb-fncn}
\psi_{\widehat{b}}(q)=\left(C_+(q)\exp(\frac{i}{h}S_+(q,b)+\frac{\pi i}{4})+C_-(q)\exp(\frac{i}{h}S_-(q,b)-\frac{\pi i}{4})\right)\sqrt{\frac{\pa p(q,b)}{\pa b}}
\end{equation}
Here $C_\pm(q)=C_\pm^{(0)}(1+\sum_{n\geq 1}a_\pm^{(n)}(q)h^n)$ are asymptotic (formal) power series.
The coefficients $a^{(n)}(q)$ are determined recursively by the equation (\ref{1d-schr}), uniquely up to
constants that can be absorbed into $C$\footnote{In the one dimensional case we have the identity $\frac{\pa p(q,b)}{\pa b}=\frac{1}{p(q,b)}$ which allows to express the WKB asymptotic in terms of $p(q,b)$.}. Here we assume that as $h\to 0$ the sequence of eigenvalues
$\widehat{b}$ converges to $b$.

The corresponding eigenvalues $\widehat{b}$ should
satisfy the Bohr-Sommerfeld(BS) condition

\[
\int_{\cL_{\widehat{b}}} \alpha=2\pi h(n+\frac{1}{2})(1+O(h))
\]
Here $\alpha=pdq$, $n\to\infty$ as $h\to 0$ such that $hn$ is finite and $\widehat{b}\to b$.
We summarized the derivation of the BS quantization condition
through the analysis of solutions near turning points $q_1, q_2$ in the appendix \ref{t-p}.

The Hamilton-Jacobi function $S(q_\pm, b)$ can be written as
\[
S_\pm(q,b)=\int_{\gamma_\pm(q,x_0)} \alpha_b
\]
Here we made a choice of a reference point $x_0=(p_0,q_0)\in \cL_b$.
Contours $\gamma_\pm$ connect the point $x_0$ with points $(\pm p(q,b),q)$ respectively,
on $\cL_b$. The Bohr-Sommerfeld condition ensures that
exponents in the semiclassical formulae do not depend on
the choice of $\gamma$ (mod $2\pi h$). The change of the reference point $x_0$
changes the overall constant.

Natural objects in the semiclassical analysis are not functions but half-densities, see for example \cite{BW}\cite{Z}\cite{GS}\cite{GS2},
and references therein. Half-densities are particularly natural to consider when no metric or volume form on  the configuration space
is specified. In this case the space of square integrable functions is not defined but the space of square integrable half-densities is still naturally defined.

Thus, instead of (\ref{wkb-fncn}) we shall consider the eigenhalfdensity
\begin{eqnarray*}
\psi_b(q)\sqrt{|dq|} & =C(\exp(\frac{i}{h}S(q_+,b)+i\frac{\pi}{4})\left(\frac{\pa p}{\pa b}\right)^{\frac{1}{2}}(1+\sum_{n>0}h^na_+^{(n)}(q,b))+ \\
&\exp(\frac{i}{h}S(q_-,b)-i\frac{\pi}{4})\left(\frac{\pa p}{\pa b}\right)^{\frac{1}{2}}(1+\sum_{n>0}h^na_-^{(n)}(q,b)))\sqrt{|dq|}
\end{eqnarray*}
where coefficients $a^{(n)}_\pm$ are as in (\ref{wkb-fncn}).

These semiclassical eigenfunctions for real values of $h$ are exponentially decaying away from $D_b=\{q|H(p,q)=b\}$, and away from this region behave as $\exp(-\frac{c(q,b)}{h})$ as $h\to 0$, where $c(q)$ is a positive function given by the Hamilton-Jacobi action. Note that to consider $\psi_b(q)$ being a half-density in both $b$ and $q$ is even more natural:
\begin{eqnarray*}
\psi_b(q)\sqrt{|dbdq|}&=C(\exp(\frac{i}{h}S(q_+,b)+i\frac{\pi}{4})\left(\frac{\pa p}{\pa b}\right)^{\frac{1}{2}}(1+O(h))+ \\
&\exp(\frac{i}{h}S(q_-,b)-i\frac{\pi}{4})\left(\frac{\pa p}{\pa b}\right)^{\frac{1}{2}}(1+O(h))\sqrt{|dbdq|}
\end{eqnarray*}
This expression should be considered as the asymptotic of the kernel of the integral operator acting from the space of halfdensities in $q$ to space of halfdensities in $b$.

\subsection{Hamiltonians which are polynomial in momentum}
Now assume that the classical Hamiltonian $H(p,q)$ is a polynomial of order $n$ in momentum $p$ with coefficients which
may depend analytically on $q$. Its quantization is an $n$-th order self-adjoint differential operator $\widehat{H}$ whose principal symbol is the classical Hamiltonian $H(p,q)$\footnote{There is the usual ambiguity
of the choice of the quantum Hamiltonian mod(h). We assume that we made such a choice. For example, one can chose
symmetric ordering of $\frac{\pa}{\pa q}$ and $q$. For more details see for example \cite{LTa}}. We want to describe its
semiclassical  eigenvectors, i.e. asymptotical behaviour as $h\to 0$ of square-integrable solutions to
\[
\widehat{H}(-ih\frac{\pa}{\pa q}, q)\psi_{\widehat{b}}(q)=\widehat{b}\psi_{\widehat{b}}(q)
\]

As in the previous section let us assume that the level curve
\begin{equation}\label{pol-schr}
\cL_b=\{(p,q)|H(p,q)=b\}
\end{equation}
of the classical Hamiltonian is connected and compact. Otherwise we will have tunneling effect between components of
WKB eigenfunctions supported on connected components of $\cL_b$, which deserves a separate discussion.
From now on we will not indicate the difference between $\widehat{b}$ and $b$, assuming that this is clear.

\begin{figure}[htb]
\includegraphics[height=6cm,width=8cm]{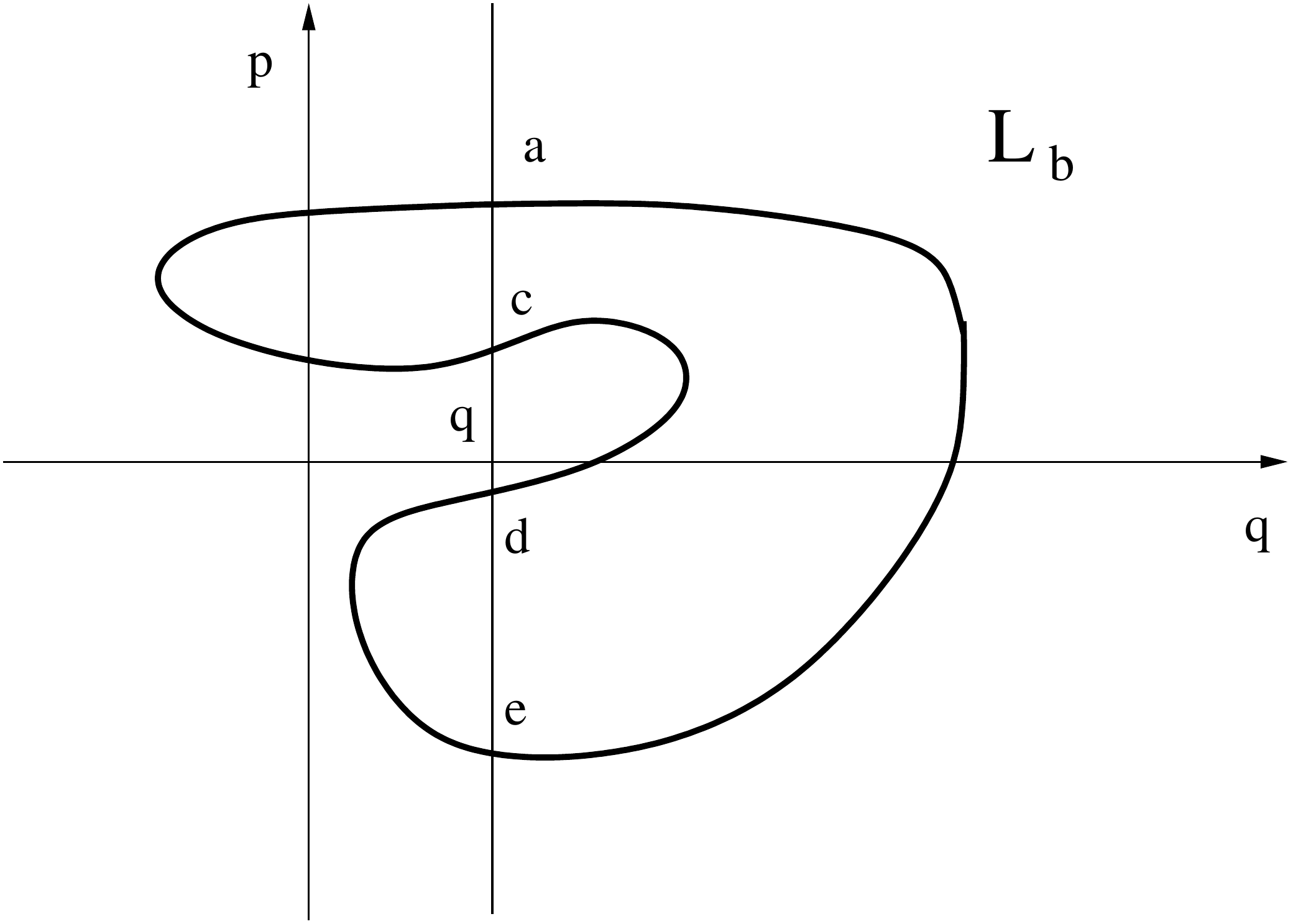}
\caption{Level curve of a Hamiltonian which is polynomial in $p$
with intersection points $a,c,d,e\in T^*_q\RR\cap \cL_b$.}
\label{F2}
\end{figure}

Normalized $L_2$ eigenhalfdensities  in the limit $h\to 0$ have the following semiclassical asymptotic\footnote{The formula (\ref{wkb-polyn}) is a linear combination of formal local eigenfunctions
of the Schrodinger operator where coefficients are fixed by the standard analysis of turning points
(see Appendix \ref{t-p} for more details).}
\begin{eqnarray}\label{wkb-polyn}
\psi_{b}(q)\sqrt{|dqdb|}=\frac{C}{\sqrt{2\pi h}}\sum_a \left|\frac{\pa p(a,b)}{\pa b}\right|^{\frac{1}{2}}& \exp(\frac{i}{h}S_\gamma
(a,b)+i\frac{\pi}{2}\mu_\gamma(a,x_0))\\&(1+\sum_{n>0}h^nc^{n}(a,b)))\sqrt{|dqdb|} \nonumber
\end{eqnarray}
Here $|C|=1$ and the sum is taken over preimages of $q$ on branches of $\cL_b$ over $q$. The coefficients $c^{n}(a)$ are uniquely (up to a constant)
determined by the equation (\ref{pol-schr}) and can be computed recursively. We fixed a reference point $x_0$ and a path $\gamma\subset \cL_b$ connecting each critical point
with the reference point. The coefficient $\mu_\gamma(a,x_0)$ is the difference between the number of negative  and positive passings through
turning points along $\gamma$ in the direction from  $x_0$ to $a$ (see appendix \ref{t-p}).
This number is also known as the Maslov index of $\gamma\subset \cL_b$.

The corresponding eigenvalue should satisfy
the  Bohr-Sommerfeld quantization condition:
\[
\int_{\cL_b} \alpha=2\pi h (n+\frac{m}{2})(1+O(h))
\]
The integral is taken in the counter-clockwise direction with respect to
the orientation on $\RR^2$ given by the symplectic form $dp\wedge dq$. Here $m$ is the Maslov index of $\cL_b$ and $m=1/2$ because all level curves are contarctible in $\RR^2$.

The Hamilton-Jacobi action function $S_\gamma(a,b)$ is determined by the property $\frac{\pa S_{\gamma}(a,b)}{\pa q}=p_a(q,b)$,
where $p_a(q,b)$ is the branch of the level curve over a neighborhood of $q$ in the vicinity of the intersection point $a$.
It is given by the integral
\[
S_\gamma(a, b)=\int_{\gamma_{a,x_0}}\alpha_b
\]
Here the form $\alpha_b$ is the restriction of $\alpha=pdq$ to $\cL_b$ and
the path $\gamma_{a,x_0}$ connects the reference point $x_0$ with $a$
on $\cL_b$. In other words $S_\gamma(a,b)$ is a generating function for $\alpha$ restricted to $\cL_b$.

Bohr-Sommerfeld conditions ensure that the combination $\frac{i}{h}S_\gamma
(a,b)+i\frac{\pi}{2}\mu_\gamma(a,x_0)$ does not depend on the choice of $\gamma$ modulo $2\pi \ZZ$.
Changing the reference point changes an overall constant $C$.

When the Hamiltonian is not polynomial but level, the curve $\cL_b$ has  finitely
many fibers over each $q$, the WKB eigenfunctions for eigenvalues in a neighborhood of $b$ are given by the same expressions.

Observing that $\frac{\pa p(a,b)}{\pa b}=\frac{\pa^2 S_\gamma(a,b)}{\pa q\pa b}$ we can write (\ref{wkb-polyn}) as
a section of the bundle of half-forms (see \cite{GS} for details) on $B\times \RR$:
\begin{equation*}
\psi_{b}(q)\sqrt{db\wedge dq}=\frac{C}{\sqrt{2\pi h}}\sum_a \exp(\frac{i}{h}S_\gamma
(a,b))(1+O(h))\sqrt{\frac{\pa^2 S_\gamma(a,b)}{\pa q\pa b}db\wedge dq}
\end{equation*}
Here $B=\{b\in \RR| H(p,q)=b, (p,q)\in T^*\RR\}$ and the branch of the square root is determined by the Maslov index $\mu_\gamma(a,x_0)$.
Note that $\frac{\pa^2 S_\gamma(a,b)}{\pa q\pa b}db\wedge dq=dp\wedge dq$.



\subsection{Two Hamiltonian systems on $T^*\RR$}
Let $H_1$ and $H_2$ be two Hamiltonian functions on $T^*\RR\simeq \RR^2$. As above we will use
coordinates $p, q$. The standard symplectic form $\omega=dp\wedge dq$ fixes an orientation on $\RR^2$.
Let $\widehat{H_1}$ and $\widehat{H_2}$ be self-adjoint differential operators
quantizing $H_1$ and $H_2$ respectively.

The goal of this section is to describe the scalar product of normalized
WKB eigen-half-densities:
\begin{multline} \label{2-wkb}
\psi^{(k)}_{b_k}(q)\sqrt{|db_kdq|}=\frac{C_k}{(2\pi h)^{1/2}}\sum_{a\in \cL_k\cap T^*_q\RR}\exp\left(\frac{i}{h}S_{\gamma_k(a,x_0^{(k)})}(a,b_k)+i\frac{\pi}{2}\mu_{\gamma_k}(a,x_0^{(k)})\right)\\ \left|\frac{\pa p^{k}_a(q,b_k)}{\pa b_k}\right|^{\frac{1}{2}}(1+O(h))\sqrt{|db_kdq|}
\end{multline}
Here $k=1,2$, $b_k\in B^{k}$ where $B^k$ is the set of all possible values of $H_k(p,q)$. Denote by $\cL_k=\{(p,q)\in \RR^2|H_k(p,q)=b_k\}$ corresponding
level curves of Hamiltonians. The sum in (\ref{2-wkb}) is taken over fibers of
$\cL_k$ over $q$ and the rest of the notations is explained in the previous section. We made choices of
reference points $x_0^{(k)}\in \cL_k$.

The scalar product of halfdensities $f=f(b_1,q)\sqrt{|db_1dq|}$ and $g=g(b_2,q)\sqrt{|db_2dq|}$ on $B_1\times \RR$ and $B_2\times \RR$ respectively
is the halfdensity on $B_1\times B_2$ given by the integral over $q$:
\[
(f,g)=\int_\RR \overline{f(b_1,q)}g(b_2,q)|dq| \sqrt{|db_1db_2|}
\]
Here we integrate the density $\overline{f}g=\overline{f(b_1,q)}g(b_2,q)|dq|$ over $\RR$.

\begin{proposition} The scalar product of generic eigen-half-densities of $\widehat{H}_1$ and
$\widehat{H}_2$  has the following semiclassical asymptotic
\begin{multline}\label{wkb2sys}
(\psi^{(2)}_{b_2},\psi^{(1)}_{b_1})=C\sum_{c} \frac{1}{(2\pi h)^{1/2}}
\exp(\frac{i}{h}S_{\gamma_1,\gamma_2}(c,b_1,b_2)+\frac{\pi i}{2}\mu_{\gamma_1,\gamma_2}(c,c_0)) \\
\left|\frac{\pa^2S_{\gamma_1,\gamma_2}(c,b_1,b_2)}{\pa b_1\pa b_2}\right|^{\frac{1}{2}}(1+O(h))\sqrt{|db_1db_2|}
\end{multline}
Here $|C|=1$, the sum is taken over intersection points of $\cL_1$ and $\cL_2$, we made a choice of reference intersection point $c_0$ and $\mu_{\gamma_1,\gamma_2}(c,x_0)$ is
the number of points along $\gamma_1$ where $\gamma_1$ is parallel to level curves of $H_2$\footnote{Up to a constant, it is also equal to minus the number of
points between $x_0$ and $c$ along $\gamma_2$, where the level curves of $H_1$ are parallel to $\gamma_2$.}
and $S_{\gamma_1,\gamma_2}(c,b_1,b_2)=S_{\gamma_1}(c,b_1)-S_{\gamma_2}(c,b_2)$.  The exponent does not depend on the choice of $\gamma_1$ and $\gamma_2$
if the Bohr-Sommerfeld quantization conditions for $b_1$ and $b_2$ hold\footnote{ As it is shown in remark \ref{genf} the exponent is the generating function for local symplectomorphism near the intersection of point of $\cL_1$ and $\cL_2$ which bring one Lagrangian fibration to the other.}. The change of $x_0$ only changes the constant $C$.

\end{proposition}

Note that the difference $S_{\gamma_1,\gamma_2}(c_1,b_1,b_2)-S_{\gamma_1,\gamma_2}(c_2,b_1,b_2)=\int_{D_{\gamma_1,\gamma_2}}\omega$ is the symplectic area of a disc which is bounded by curves $\gamma_1$ with its own orientation
and by $\gamma_2$ taken with the opposite orientation between two intersection points $c_1$ and $c_2$. Because (\ref{wkb2sys})
is defined up to an arbitrary overall constant $C$, only such differences are important.

Here is an outline of the proof. We should evaluate the asymptotic of the integral
\begin{align}\label{sprod}
(\psi^{(2)}_{b_2},\psi^{(1)}_{b_1})&=\frac{C_1\overline{C_2}}{2\pi h}\sum_{a,c}\int_{q\in \RR} \exp\left(\frac{i}{h}S_{\gamma_1}(a,b_1)-S_{\gamma_2}(c,b_2)+i\frac{\pi}{2}\mu_{\gamma_1}(a,x_0^{(1)})-i\frac{\pi}{2}\mu_{\gamma_2}(c,x_0^{(2)})\right)\\
&\left|{\frac{\pa p_1(a,b_1)}{\pa b_1}\frac{\pa p_2(c,b_2)}{\pa b_2}}\right|^{\frac{1}{2}}(1+O(h))|dq| \sqrt{|db_1db_2|}
\end{align}
by the stationary phase method. Here $a\in T^*_q\RR\cap \cL_{b_1}^{(1)}$ and $c\in T^*_q\RR\cap \cL_{b_2}^{(2)}$.

Critical points of the exponent are solutions to the
equation
\[
\frac{\pa S_{\gamma_1}(a,b_1)}{\pa q}=\frac{\pa S_{\gamma_2}(c,b_2)}{\pa q}
\]
which is equivalent to
\[
p_1(a,b_1)=p_2(c,b_2)
\]
where $(p^{(1)}(a,b_1), a)$ and $(p^{(2)}(c,b_2), c)$ are
fibers over $q$ in $\cL_1$ and $\cL_2$ respectively.
Therefore, critical points of $S_{\gamma_1}(a,b_1)-S_{\gamma_2}(c,b_2)$ occur only when $a=c$ and they are intersection points
of $\cL_1$ and $\cL_2$.

Differentiating the equation $p_1(c,b_1)=p_2(c,b_2)$ we have

\[
\frac{\pa c}{\pa b_2}\frac{\pa p_2}{\pa q}(c,b_2)+\frac{\pa p_2}{\pa b_2}(c,b_2)=\frac{\pa c}{\pa b_2}\frac{\pa p_1}{\pa q}(c,b_1)
\]

For the second derivative of $S_{\gamma_1,\gamma_2}$ we have
\[
\frac{\pa^2 S_{\gamma_1,\gamma_2}(c,b_1,b_2)}{\pa b_1\pa b_2}=\frac{\pa}{\pa b_2}\left( \int_{\gamma_1}\frac{\pa p_1}{\pa b_1} dq\right)=\frac{\pa c}{\pa b_2}\frac{\pa p_1}{\pa q}(c,b_1)
\]
Combining these formulae we have the identity
\begin{equation}
\frac{\pa^2 S_{\gamma_1,\gamma_2}(c,b_1,b_2)}{\pa b_1\pa b_2}=\left(\frac{\pa p_1}{\pa q}(c,b_1)-\frac{\pa p_2}{\pa q}(c,b_2)\right)^{-1}\frac{\pa p_1}{\pa b_1}(c,b_1)
\frac{\pa p_2}{\pa b_1}(c,b_2)
\end{equation}

Now we can compute the asymptotic of the integral by the stationary phase method:
\begin{align*}\numberthis\label{2sys-proof-wkb}
(\psi^{(2)}_{b_2},\psi^{(1)}_{b_1})&=\sum_{c} \frac{C_1\overline{C_2}}{(2\pi h)^{n/2}}\exp{(\frac{\pi i}{4}\sign(\frac{\pa^2 S_{\gamma_1,\gamma_2}}{\pa q^2}(c,b_1,b_2)+i\frac{\pi}{2}\mu_{\gamma_1}(a,x_0^{(1)})-i\frac{\pi}{2}\mu_{\gamma_2}(c,x_0^{(2)}))}\\
&\exp{\frac{i}{h}(S_{\gamma_1}(c,b_1)-S_{\gamma_2}(c,b_2))}
\left|\frac{\frac{\pa p^{(1)}}{\pa b_1}(c,b_1)\frac{\pa p^{(2)}}{\pa b_2}(c,b_2)}{\frac{\pa^2 S_{\gamma_1,\gamma_2}}{\pa q^2}(c, b_1,b_2)}\right|^{\frac{1}{2}} (1+O(h))\sqrt{|db_1db_2|}
\end{align*}
One can show $\mu_{\gamma_1}(c,x_0^{(1)})-\mu_{\gamma_2}(c,x_0^{(2)})+\frac{1}{2} sign(\frac{\pa^2 S_{\gamma_1,\gamma_2}}{\pa q^2}(c,b_1,b_2)=\mu_{\gamma_1,\gamma_2}(c,c_0)+A(c_0,x^{(1)},x^{(2)})$. The constant $A(c_0,x^{(1)},x^{(2)})$ does not depend on $c$ and is absorbed into an overall constant. Taking into account that $\frac{\pa^2 S_{\gamma_1,\gamma_2}(c,b_1,b_2)}{\pa q^2}=\frac{\pa p_1}{\pa q}(c,b_1)-\frac{\pa p_2}{\pa q}(c,b_2)$ this gives the formula (\ref{wkb2sys}).

Note that (\ref{wkb2sys}) can also be written as section of the bundle of half-forms \cite{GS} over $B_1\times B_2$
\[
(\psi^{(2)}_{b_2},\psi^{(1)}_{b_1})=\sum_{c} \frac{C}{(2\pi h)^{1/2}}
\exp(\frac{i}{h}S_{\gamma_1,\gamma_2}(c,b_1,b_2))
(1+O(h))\sqrt{\frac{\pa^2S_{\gamma_1,\gamma_2}(c,b_1,b_2)}{\pa b_1\pa b_2}db_1\wedge db_2}
\]
Here the branch of the square root is determined by $\mu_{\gamma_1,\gamma_2}(c,c_0)$. Let $H_1^*\times H_2^*: \Omega^\bullet(B_1\times B_2)\to \Omega^\bullet(\RR^2)$
be the pullback of the projection to the space of level curves of $H_1$ and $H_2$. Then it is easy to check that at a smooth point
\[
H_1^*\times H_2^*(\frac{\pa^2S_{\gamma_1,\gamma_2}(c,b_1,b_2)}{\pa b_1\pa b_2}db_1\wedge db_2)=\omega(c)
\]

\begin{remark} It is also easy to see that
\[
\frac{\pa^2 S_{\gamma_1,\gamma_2}}{\pa b_1\pa b_2}(c,b_1,b_2)=\{H_1, H_2\}(c)^{-1}
\]
\end{remark}

\begin{remark}
Scalar products $(\psi^{(2)}_{b_2},\psi^{(1)}_{b_1})$ can be regarded as eigenfunctions of
$\widehat{H}_1$ when this operator is written in the basis of eigenvectors of $\widehat{H}_2$ (or vice versa).
If level curves of the second Hamiltonian are not compact, there is no Bohr-Sommerfeld quantization condition.
When  $H_2=p$ this formula reduces to
WKB eigenfunctions (\ref{wkb-polyn}).
\end{remark}

\section{Semiclassical eigenfunctions for an integrable systems on a cotangent bundle}\label{ndwkb}

\subsection{An integrable system on a cotangent bundle}

Recall that the cotangent  bundle  $T^*Q_n$ to a smooth manifold $Q_n$ has the natural symplectic
structure which in local coordinates\footnote{Here $q^i$ are local coordinates on $Q_n$ and $p_i$ are corresponding coordinates on the fiber.} is
\[
\omega=\sum_{i=1}^n dp_i\wedge dq^i
\]
This symplectic form is exact $\omega=d\alpha$ where $\alpha$ is a globally defined $1$-form which in
local coordinates is $\alpha=\sum_{i=1}^n p_i dq^i$.

The bundle projection $\pi: T^*Q_n \to Q_n$  is a Lagrangian fibration. The mapping $q\mapsto (0,q)$ is a section of this projection
known as the zero-section. It gives an embedding of $Q_n\subset T^*Q_n$ as a Lagrangian submanifold.

Geometrically, an integrable system on a symplectic manifold $\cM_{2n}$ is a
Lagrangian fibration. Algebraically this is a choice of a maximal Poisson commutative subalgebra in the algebra of
functions on $T^*Q$. For example, it can be generated by $N$ Poisson commuting functions
$f_1, \dots, f_N$ of which only $n$ are linearly independent. The base of the fibration in
this case is a variety $B$ defined by equations $P_1(f_1,\dots, f_N)=P_2(f_1,\dots, f_N)=\dots=0$,
where $P_1,\dots P_{N-n}$ are polynomials. The Lagrangian fibers in this case are level surfaces of $f_1,\dots, f_N$.

For simplicity assume that we have $n$ independent functions $H^1, \dots, H^n$ on $\cM_{2n}$
which Poisson commute. This defines Lagrangian fibration $H: \cM_{2n}\to B_N\subset \RR^n$, $x\mapsto (H^1(x), \dots, H^n(x))$ given by level surfaces of Hamiltonians $\{H^i\}$.  Denote by $\cL_b$ the level surface
\[
\cL_b=\{x\in \cM_{2n}| H^i(x)=b^i\}
\]
i.e. the fiber of $H$ over $b$.

For generic $b$\footnote{For special $b$ it may degenerate and there is an interesting analysis of monodromies related to singularities, see for example \cite{N}\cite{BCD} and references therein, but we will ignore this here.} the fiber $\cL_b$ is a Lagrangian submanifold. By the Liouville theorem such level surfaces are isomorphic to $T^k\times \RR^{n-k}$
for some $k$, where $T^k$ is a $k$-dimensional torus.

Let $(p, q)$ be local Darboux coordinates on $\cM_{2n}=T^*Q_n$, such that $q^i$ are local coordinates on $Q_n$ and $p_i$ are
coordinates on fibers. Let $c=(p_c(q,b),q)$ be the point on $\cL_b$ which projects on $q$. We will use abbreviated notation
$p(c,b)$ instead of $p_c(q,b)$. Then on a tubular neighborhood of $\cL_b$
the symplectic form can be written as\footnote{The angle variables $\varphi_i$ are affine coordinates on level surfaces $\cL_b$ defined by the Hamiltonian
flows generated by Hamiltonians $H_i$. They form a complementary set of coordinates to $b^i$ in
a Darboux coordinate chart covering a neighborhood of $t$'s branch of $\cL_b$ and :
\[
d\varphi_i=\sum_{i=1}^n  \frac{\pa p_i(c,b)}{\pa b^j}  dq^j
\]
So that $\omega=\sum_{i=1}^n db^i\wedge d\varphi_i$.}
\[
\omega=\sum_{i=1}^n dp_i\wedge dq^i=\sum_{i,j=1}^n \frac{\pa p_i(c,b)}{\pa b^j}db^j\wedge dq^i
\]

The pull-back of the one form
\[
\alpha=\sum_{i=1}^n p_i dq^i
\]
 to $\cL_b$ is closed because $\cL_b$ is Lagrangian and $d\alpha=\omega$. Therefore, for a closed curve $C\subset \cL_b$ the
 integral $\int_C\alpha$ does not depend on continuous deformations of $C$ and is the pairing of homology class of $C$ with the de Rham
 cohomology class of $\alpha$.

\subsection{Quantization of the cotangent bundle}
\subsubsection{}The algebra of differential operators on a smooth $n$-dimensional manifold $Q$ forms a natural deformation quantization of the
Poisson algebra $\cO(T^*Q)$ of functions which are smooth on $Q$ and polynomial in cotangent directions.

Denote by $D_h(Q)$ the algebra of differential operators on $Q$ which locally have the form:
\[
D=\sum_{k_1,\dots, k_n\geq 0} h^{k_1+\dots +k_n} v_{k_1,\dots, k_n}(q)\pa_1^{k_1}\dots \pa_n^{k_n}
\]
where $\pa_l=\frac{\pa}{\pa q^l}$ and $v_{k_1,\dots, k_n}(q)$ are some smooth functions. The space $D_h(Q)$ is filtered by the degree of differential
operators. Its associate graded space is naturally isomorphic to $\cO(T^*Q)$ with the grading being the degree
of the polynomial in the cotangent direction. The multiplication of differential operators becomes
the pointwise multiplication in $\cO(T^*Q)$ and the commutator becomes a Poisson bracket. This can be also written as
\[
D_1D_2=[D_1][D_2]+O(h), \ \  D_1D_2-D_2D_1=h\{[D_1],[D_2]\}+O(h^2)
\]
where $[D]\in \cO(T^*Q)$ is the symbol of the differential operator $D$.

\subsubsection{} Let $f_1, \dots, f_N$ be Poisson commuting functions on $\cO(T^*Q)$
defining an integrable system (with only $n$ of them independent). A quantization of this
integrable system is a collection of $N$ Hermitian commuting differential operators $\widehat{f}_1, \dots \widehat{f}_N$ of
which only $n$ are independent. For simplicity assume that $Q$ is compact and that we have $n$ independent commuting differential
operators $\widehat{H}_1, \dots, \widehat{H}_n$.

\subsection{Semiclassical eigenfunction for integrable systems on $T^*Q_n$}

Assume we have an integrable system on $T^*Q_n$, i.e. a Lagrangian fibration given
by level curves of $n$ Poisson commuting independent functions (as above).

From now on to avoid the
discussion of tunneling effects we will assume that level surfaces $\cL_b$ are connected.

The projection $\pi: \cL_b\to Q_n$ is a branch cover over sufficiently small neighborhood for generic $q$.
Choose a reference point $x_0=(p_0,q_0)\in \cL_b$. Let $a\in \cL_b$ be a point in the fiber over $q$.
Define the action function $S_\gamma(a,b)$ as a generating function for the form $\alpha$ restricted to $\cL_b$,
i.e. $dS_\gamma(a,b)=\alpha$. It can be chosen as
\begin{equation}\label{act-scl}
S_\gamma(a,b)=\int_{\gamma_{a,x_0}} \alpha
\end{equation}
Here the integration is taken along a path $\gamma_{a,x_0}$ in $\cL_b$, connecting the point $a$ and the reference point $x_0\in \cL_b$.

The semiclassical asymptotic of normalized joint eigenfunctions (eigen-half-densities) of commuting quantum Hamiltonians $\widehat{H}^i, \ i=1,\dots, n$
\begin{equation}\label{Schr-eq}
\hat{H}^i\psi_b(q)=b^i\psi_b(q)
\end{equation}
is well known and is
\begin{align*}\numberthis\label{wkbtq}
\psi_b(q)=\frac{C}{(2\pi h)^{\frac{n}{2}}}\sum_c \exp(\frac{i}{h}S_\gamma(c,b)+i\frac{\pi}{2}\mu_\gamma(c,x_0))\\
\left|\det\left(\frac{\pa p_i(c,b)}{\pa b_j}\right)\right|^{\frac{1}{2}}(1+\sum_{n>0}h^na_n(c,b))\sqrt{|dbdq|} ,
\end{align*}
Here $|C|=1$, the sum is taken over the fibers of the projection $\pi: \cL_b\to Q, \pi(p,q)=q$ over $q$, coefficients $a_n(c,b)$ can be computed recursively from (\ref{Schr-eq}) and $\sqrt{dbdq}$
is the Euclidean 1/2-density in local coordinates on $B\times Q_n$. The 1/2-density
\[
\left|\det\left(\frac{\pa p_i(c,b)}{\pa b_j}\right)\right|^{\frac{1}{2}}\sqrt{|dbdq|}
\]
is defined globally on $B\times Q$ and in particular does not depend on the choice of local coordinates.
In (\ref{wkbtq}) we choose a base point $x_0\in \cL_b$ and the path $\gamma$ as in the discussion above.
The number $\mu_\gamma(c,x_0)$ is the oriented number of points on path $\gamma$ at which the tangent line (in $T(T^*Q)$)
lies  in a tangent space to a fiber of $T^*Q\to Q$, i.e. to $T_p(T^*_qQ)\subset T_{(p,q)}(T^*Q)$\footnote{Such points $a$ are higher dimensional versions
of turning points.}. Changing $x_0$ will change the overall constant $C$.

The exponent does not depend on the choice of $\gamma$ if Bohr-Sommerfeld quantization conditions on $b$
\[
\int_\beta\alpha_b=2\pi h (n_\beta+\frac{m(\beta)}{2})(1+O(h))
\]
for each non contructible cycle $\beta\subset \cL_b$ with $n_\beta\in \ZZ$ . Here $m(\beta)$ is the Maslov index of $\beta$.

Note that because
\[
\frac{\pa p_i(c,b)}{\pa b_j}=\frac{\pa^2 S_{\gamma}(c,b)}{\pa q^i\pa b^j}
\]
the asymptotical eigenhalfdensity (\ref{wkbtq}) defines the following half-form on $B\times Q$:
\[
\psi_b(q)=\frac{C}{(2\pi h)^{\frac{n}{2}}}\sum_c \exp(\frac{i}{h}S_\gamma(c,b))(1+\sum_{n>0}h^na_n(c,b))\sqrt{\left(\frac{\pa^2 S_{\gamma}(c,b)}{\pa q^i\pa b^j} db^j\wedge dq^i\right)} ,
\]
The index $\mu_\gamma(c,x_0)$
indicates which branch of the square root should be taken. Also note that $(H^*\times \pi^*)(\frac{\pa^2 S_{\gamma}(c,b)}{\pa q^i\pa b^j} db^j\wedge dq^i)=\omega(c)$
where $H^*$ is the pullback on forms of the projection $H\times \pi: T^*Q\to B\times Q$.

Though both, the half-density and the half-form above, are written in local coordinates, they
are globally defined.


\subsection{Two integrable systems} Let $H_1$ and $H_2$ be corresponding Lagrangian fibrations by level surfaces of
integrals of two integrable systems on $T^*Q$
\begin{equation}\label{HH}
H_{1}: T^*Q\to B_{1}, \ \ H_{2}: T^*Q \to B_{2}
\end{equation}
Denote their fibers by $\cL^{(1)}_{b_1}$ and $\cL^{(2)}_{b_2}$ respectively. We assume that $H_1$ and $H_2$
are transversal Lagrangian fibrations, i.e. their generic fibers intersect transversally and, in particular, over finitely
many points.

The formula for the semiclassical asymptotic of the scalar product of two
WKB eigen-halfdensities corresponding $\cL^{(1)}_{b_1}$ and $\cL^{(2)}_{b_2}$ can be derived similarly to the one-dimensional case
discussed in the previous section.
\begin{multline}\label{wkb2systq}
(\psi^{(2)}_{b_2},\psi^{(1)}_{b_1})= \frac{C}{(2\pi h)^{n/2}}\sum_{c}
\exp(\frac{i}{h}S_{\gamma_1,\gamma_2}(c,b_1,b_2))+\frac{\pi i}{2}\mu_{\gamma_1,\gamma_2}(c,x_0)) \\
\left|\det\left(\frac{\pa^2 S_{\gamma_1,\gamma_2}(c,b_1,b_2)}{\pa b_1^i\pa b_2^j}\right)\right|^{\frac{1}{2}}\sqrt{|db_1db_2|}(1+O(h))
\end{multline}
Here the sum is taken over intersection points of $\cL_{b_1}^{(1)}$ and $\cL_{b_2}^{(2)}$ and $\sqrt{|db_1db_2|}$
is the Euclidean half-density on a  coordinate neighborhood in $B_1\times B_2$. We choose reference points
$x_0^{(i)} \in \cL_{b_i}^{(i)}$ and two paths, $\gamma_1$ connecting $x_0^{(1)}$ and $c$ in $\cL^{(1)}$ and $\gamma_2$,
connecting $x_0^{(2)}$ and $c$ in $\cL^{(2)}$. The function $S_{\gamma_1,\gamma_2}(c,b_1,b_2))$ is defined as
\[
S_{\gamma_1,\gamma_2}(c,b_1,b_2)=S_{\gamma_1}^{(1)}(c,b_1)-S_{\gamma_2}^{(2)}(c,b_2)
\]
where $S_{\gamma_k}(c,b_k)$ are as in (\ref{wkbtq}). Note that if $\gamma_1$ and $\gamma_2$ pass through two intersection points $a$ and $b$ we have
\[
S_{\gamma_1,\gamma_2}(c_1,b_1,b_2)-S_{\gamma_1,\gamma_2}(c_2,b_1,b_2)=\int_{D_{\gamma_1,\gamma_2}} \omega
\]
where $D_{\gamma_1,\gamma_2}$ is a disk with the boundary formed by segments of $\gamma_1$ and $\gamma_2$
between two intersection points $c_1$ and $c_2$. If Bohr-Sommerfeld quantization conditions hold, then, modulo $2\pi h \ZZ$, this difference does not depend on paths.
The integer $\mu_{\gamma_1,\gamma_2}(c,x_0)$ is the number of times when the tangent line to $\gamma_1$
is inside the tangent plane to a level surface of $H^{(2)}$\footnote{ As in the one dimensional case,
this number, the Maslov index, is also also equal to minus the number of times when the tangent plane to $\gamma_2$ is inside
the tangent plane of a level surface of $H^{(1)}$.}.

The exponent in (\ref{wkb2systq}) does not depend on choices of $\gamma_1$ and $\gamma_2$ if Bohr-Sommerfeld
conditions hold for $\cL^{(1)}$ and for $\cL^{(2)}$.

Finally, the half-density (\ref{wkb2systq}) is globally defined. It is easy to see that it does
not depend on local coordinates on $B_1\times B_2$. Moreover
\[
H_1^*\times H_2^*(\frac{\pa^2 S_{\gamma_1,\gamma_2}(c,b_1,b_2)}{\pa b_1^i\pa b_2^j}db_1^i\wedge db^j)=\omega(c)
\]
where $H_1\times H_2: T^*Q\to B_1\times B_2$ is the product of the  Lagrangian projections (\ref{HH}).

The proof of the formula (\ref{wkb2systq}) is similar to the one dimensional case. We should evaluate the asymptotic of the integral
\begin{align*}
(\psi^{(2)}_{b_2},\psi^{(1)}_{b_1})&=\frac{C_1\overline{C_2}}{(2\pi h)^{n}}\sum_{s,t}\int \exp\left(\frac{i}{h}\left(S_{\gamma_1}(a,x_0)-S_{\gamma_2}(c,x_0)\right)+i\frac{\pi}{2}\mu^{(1)}_{\gamma_1}(a,x_0)-i\frac{\pi}{2}\mu^{(2)}_{\gamma_2}(c,x_0)\right)\\
&\left|\det\left(\frac{\pa p^{(1)}(a,b_1)}{\pa b_1}\right)\det\left(\frac{\pa p^{(2)}(c,b_2)}{\pa b_2}\right)\right|^{\frac{1}{2}}(1+O(h))|dq|\sqrt{|db_1db_2|}
\end{align*}
as $h\to 0$ by the stationary phase method. Critical points of the exponent are solutions to the
equation
\[
d_qS_{\gamma_1}^{(1)}(a,b_1)=d_qS_{\gamma_2}^{(2)}(c,b_2)
\]
which is exactly the equation $\alpha_{b_1}(a)=\alpha_{b_2}(c)$. Its solutions are intersection points of two Lagrangian sumbanifolds $\cL^{(1)}_{b_1}\cap \cL^{(2)}_{b_2}$.
Applying the stationary phase method to these critical points we have
\begin{multline*}\label{2sys-proof-wkb}
(\psi^{(2)}_{b_2},\psi^{(1)}_{b_1})=\frac{C_1\overline{C_2}}{(2\pi h)^{n/2}}\sum_{c\in \cL^{(1)}_{b_1}\cap\cL^{(2)}_{b_2}} \exp(\frac{i}{h}(S^{(1)}_{\gamma_1}(c,b_1)-S^{(2)}_{\gamma_2}(c,b_2))+i\frac{\pi}{2}\mu^{(1)}_{\gamma_1}(a,x_0)-i\frac{\pi}{2}\mu^{(2)}_{\gamma_2}(c,x_0)+\\
i\frac{\pi}{4}\sign(B(c,b_1,b_2)))
\left|\frac{\det(\frac{\pa^2 S_{\gamma}(c,b_1)}{\pa q\pa b_1})\det(\frac{\pa^2 S_{\gamma}(c,b_2)}{\pa q\pa b_2})}{\det(B(c,b_1,b_2))}\right|^{\frac{1}{2}} (1+O(h))\sqrt{|db_1db_2|}
\end{multline*}
where
\[
B(c,b_1,b_2)_{i,j}=\frac{\pa^2 S_{\gamma_1}^{(1)}}{\pa q^i\pa q^j}(c, x_0)-\frac{\pa^2 S_{\gamma_2}^{(2)}}{\pa q^i\pa q^j}(c, x_0)=
\frac{\pa p_i^{(2)}}{\pa q^j}(c,b_1)-\frac{\pa p_i^{(1)}}{\pa q^j}(c,b_2)
\]

For intersection points of $\cL_{b_1}^{(1)}$ and $\cL_{b_2}^{(2)}$ we have $p^{(1)}(c, b_1)=p^{(2)}(c, b_2)$. Differentiationg this identity in $b_2$ we
obtain:
\[
\sum_k\frac{\pa c^k}{\pa b^j_2}\frac{\pa p^{(1)}_i(c,b_1)}{\pa c^k}=\sum_k\frac{\pa c^k}{\pa b^j_2}\frac{\pa p^{(2)}_i(c,b_2)}{\pa c^k}+\frac{\pa p^{(2)}_i(c,b_2)}{\pa b^j_2}
\]
On the other hand
\begin{equation}\label{sder}
\frac{\pa^2 S_{\gamma_1, \gamma_2}}{\pa b_1^i\pa b_2^j}(c, x_0)=\frac{\pa}{\pa b^j_2} \int_{\gamma_1} \sum_k\frac{\pa p^{(1)}_i(q,b_1)}{\pa q^k}dq^k=\sum_k\frac{\pa c^k}{\pa b^j_2}\frac{\pa p^{(1)}_i(c,b_1)}{\pa b_1^k}
\end{equation}
The first identity implies
\[
\frac{\pa c}{\pa b_2}=(\frac{\pa p^{(1)}(c,b_1)}{\pa c}-\frac{\pa p^{(2)}(c,b_2)}{\pa c})^{-1}\frac{\pa p^{(2)}(c,b_2)}{\pa b_2}
\]
Substituting this into (\ref{sder}) we obtain
\begin{equation}\label{eq-1}
\frac{\pa^2 S_{\gamma_1, \gamma_2}}{\pa b_1\pa b_2}(c, x_0)=\frac{\pa p^{(1)}(c,b_1)}{\pa b_1}\left(\frac{\pa p^{(1)}(c,b_1)}{\pa c}-\frac{\pa p^{(2)}(c,b_2)}{\pa c}\right)^{-1}\frac{\pa p^{(2)}(c,b_2)}{\pa b_2}
\end{equation}

The complete proof, with analytical details, and with the details on Maslov indices will be given elsewhere.

\begin{remark} Let us show that
\begin{equation}\label{SH}
\frac{\pa^2 S_{\gamma_1, \gamma_2}}{\pa b_1\pa b_2}(c, x_0)=(\{H_1, H_2\})^{-1}
\end{equation}
where $(\{H_1, H_2\})_{ij}=\{H_1^i, H_2^j\}$ is the matrix of Poisson brackets.

Indeed, denote $A_{ki}=\frac{\pa p_k(c, b_1)}{\pa b_1^i}$ and
$B_{ki}=\frac{\pa p_k(c, b_{2})}{\pa b_2^i}$.
We have :
\begin{align*}
\{H_1^i, H_2^j\}(c)&=\frac{\pa H^i_1}{\pa p_k}(c)\frac{\pa H_2^j}{\pa q^k}(c)-\frac{\pa H_1^i}{\pa q^k}(c)\frac{\pa H_2^j}{\pa p_k}(c)\\ &=
( A^{-1})^{ik}\frac{\pa H_2^j}{\pa q^k}(c)-(B^{-1})^{jk}\frac{\pa H_1^i}{\pa q^k}(c)= \\
( A^{-1})^{ik}&( B^{-1})^{jl}\left(\frac{\pa p^{(2)}_l(c, b_2)}{\pa b_{2}^m}\frac{\pa H_2^m}{\pa q^k}-
\frac{\pa p^{(1)}_k(c, b_1)}{\pa b_{1}^m}\frac{\pa H_1^m}{\pa q^l}\right)= \\
(A^{-1})^{ik}&( B^{-1})^{jl}\left(\frac{\pa p^{(2)}_l(c,b_2)}{\pa{q^k}}-
\frac{\pa p^{(1)}_k(c,b_1)}{\pa{q^l}}\right)
\end{align*}
Here $(A^{-1})^{ik}=\frac{\pa H_1^i}{\pa p_k}$ and $(B^{-1})^{ik}=\frac{\pa H_2^i}{\pa p_k}$.
Now, take into account that the form $\alpha=\sum_k p_k(c,b)dq^k$ is closed (since $\cL_b$ is
a Lagrangian submanifold). Therefore $\frac{\pa p^{(1)}_k(c,b_1)}{\pa{q^l}}=\frac{\pa p^{(1)}_l(c,b_1)}{\pa{q^k}}$.
Together with the formula above this gives
\[
\{H_1^i, H_2^j\}(c)=(A^{-1})^{ik}( B^{-1})^{jl}\left(\frac{\pa p^{(2)}_l(c,b_2)}{\pa{q^k}}-
\frac{\pa p^{(1)}_l(c,b_1)}{\pa{q^k}}\right)
\]
Combining this formula with (\ref{eq-1}) we obtain (\ref{SH}).

\end{remark}

\begin{remark}\label{genf}

Let $c\in \cL^{(1)}_{b_1}\cap \cL^{(2)}_{b_2}$ and $U_1\subset B_1, U_2\subset B_2$ be open neighborhoods of
$\pi_1(c)$ and $\pi_2(c)$ respectively. Then we have natural symplectomorphisms $\pi_1^{-1}(U_1)\simeq W_1\subset T^*U_1$ and
$\pi_2^{-1}(U_2)\simeq W_2\subset T^*U_2$. Let $\phi^{(1)}$ are affine coordinates (the angle variables) on fibers of $M\to B_1$ generated
by coordinates $b_1$ and $\phi^{(2)}$ are the angle variable corresponding to coordinates $b_2$ on $B_2$. In these coordinates
\[
\omega=db_1\wedge d\phi^{(1)}=db_2\wedge d\phi^{(2)}
\]
Let $\varphi: W_1\to W_2$ be the natural symplectomorphism mapping $(b_1,\phi^{(1)})\mapsto (b_2, \phi^{(2)})$.

\begin{theorem} The function $S_{\gamma_1,\gamma_2}(c, b_1,b_2)$ is the generating function of this symplectomorphism.
\end{theorem}

Indeed, we need to prove that
\[
\phi^{(1)}_i=\frac{\pa S_{\gamma_1,\gamma_2}(c, b_1,b_2)}{\pa b_1^i}, \ \ \phi^{(2)}_i=-\frac{\pa S_{\gamma_1,\gamma_2}(c, b_1,b_2)}{\pa b_2^i}
\]
Assume $M=T^*Q$ (if not, choose appropriate local Darboux coordinates) and $\alpha=p dq$.
Let $(p_c,q_c)$ be coordinates of $c\in \cL^{(1)}_{b_1}\cap \cL^{(2)}_{b_2}$, Then
\[
\frac{\pa S_{\gamma_1,\gamma_2}(c, b_1,b_2)}{\pa b_1^i}=\frac{\pa q_c^j}{b_1^i}(p^{(1)}_c(b_1,q_c)_j-p^{(2)}_c(b_2,q_c)_j)+ \int^{q_c}\frac{\pa p^{(1)}_c(b_1,q)_j}{\pa b_1^j}=
\int^{\phi_c}d\phi_i^{(1)}=\phi^{(1)}_i
\]
Here $\{\phi^{(1)}_i\}$ are angle coordinates on $\cL_{b_1}^{(1)}$ corresponding to coordinates $\{b_i\}$ on $B_1$.
Here we used the fact that  $c\in \cL^{(1)}_{b_1}\cap \cL^{(2)}_{b_2}$ and therefore $p_c^{(1)}(b_1,q_c)=p_c^{(2)}(b_2,q_c)$.
Similarly

\[
\frac{\pa S_{\gamma_1,\gamma_2}(c, b_1,b_2)}{\pa b_2^i}=-\int^{q_c}\frac{\pa p^{(2)}_c(b_2,q)_j}{\pa b_2^j}=\phi^{(2)}_i
\]
This proves the theorem.
\end{remark}

\section{Conclusion: some open problems and conjectures}

\subsection{General symplectic manifolds} Details of proofs and analytical aspects of the semiclassical asymptotic of eigen-half-densities will be given in a separate publication. In this concluding section
we will give some related conjectures and observations.

\subsubsection{} Let $(M, \omega)$ be a symplectic manifold. Fix geometric quantization data
which consist of the following:

\begin{itemize}
\item A line bundle $L$ (a prequantization line bundle) with Hermitian structure
on fibers and a Hermitian connection $\alpha$ on $L$ such that the symplectic
form $\omega$ is the curvature of $\alpha$, i.e. $d\alpha=\omega$ and the Hermitian
product is covariantly constant.

\item A real polarization $P\subset TM$ which is an integrable
tangent distribution on $M$ such that each generic leaf is a Lagrangian submanifold
in $M$. We assume that the space of leaves $B=M/P$ is almost everywhere smooth,
i.e. that the polarization is a Lagrangian fibration $\pi: M\to B$ (generic fibers are Lagrangian).

\end{itemize}

The space of geometric quantization $H^{(1/2)}_P$ is
the space of half-densities on $M$ which are covariantly constant (with respect to the connection $\alpha$)
along $P$. Locally it can be identified with functions on $M/P$.

Let $C_h(M)$ be quantized algebra of functions on $M$. Assume it acts on the space $H^{(1/2)}_P$.
See \cite{Ts} for the microlocal setting where $h$ is a formal variable.

\subsubsection{} A classical integrable system on $M$ is a Lagrangian fibration\footnote{The generic fibers are Lagrangian
submanifolds.} $\pi: M\to B$ which defines Poisson
commuting  subalgebra $C(M,B)\subset C(M)$ in the algebra of functions on $M$,  $C(M,B)=\pi^*(C(B))$.

Assume that the subalgebra $C(M,B)$ is quantized, i.e. deformed into a maximal commutative subalgebra $C_h(M,B)\subset C_h(M)$. For $2n$-dimensional $M$ such subalgebra has rank $n$.

Now, assume that we have a real polarization $P$ with the corresponding Lagrangian fibration $\pi_1: M\to B_1=M/P$
and an integrable system corresponding to the Lagrangian fibration $\pi_2: M\to B_2$. Assume that fibers of both projections are generically transverse. Assume that the algebra $C_h(M)$ acts on the space $H^{(1/2)}_P$.

We will say a vector $\psi_\chi\in H^{(1/2)}_P$ is an eigenvector of $C_h(M,B_2)$
corresponding to the character $\chi: C_h(M,B)\to \RR$ if
$a\psi_\chi=\chi(a)\psi_\chi$ for any $a\in C_h(M,B_2)$. Semiclassically, as $h\to 0$ the set of
characters can be identified with $B_2$.

\begin{conjecture} Semiclassical asymptotic of an eigen-half-density of $C_h(M,B_2)$ in the space $H^{(1/2)}_P$
have the following structure:

\begin{equation}\label{general}
\psi(b_1)_{b_2}=\frac{C}{(2\pi h)^{n/2}}\sum_{c\in \cL_{b_1}^{(1)}\cap \cL_{b_2}^{(2)}} e^{\frac{i}{h}S_{\gamma_1,\gamma_2}(c,b_1,b_2)+\frac{i\pi}{2}\mu_{\gamma_1,\gamma_2}(c)}\left|det\left(\frac{\pa^2 S_\gamma(c,b_1,b_2)}{\pa b_1^i\pa b_2^j}\right)\right|^{1/2}\sqrt{|db_1db_2|}
\end{equation}

Here as in (\ref{wkb2systq}) we made a choice of reference points $x_0^{(i)}\in \cL_{b_i}^{(i)}$,  $S_{\gamma_1,\gamma_2}=S^{(1)}_{\gamma_1}-S^{(2)}_{\gamma_2}$ where $S^{(i)}_\gamma=\int_{\gamma\subset \cL^{(i)}}\alpha$ and $\alpha$ is the prequantization connection, $\gamma_1$ and $\gamma_2$ are
paths in $\cL^{(1)}_{b_1}$ and $\cL^{(2)}_{b_2}$ respectively, connecting corresponding reference points and $c$.
Lagrangian submanifolds $\cL^{(1)}_{b_1}$ and  $\cL^{(2)}_{b_2}$ are fibers over $b_1\in B_1$ and $b_2\in B_2$ respectively. We assume that $\cL_{b_1}^{(1)}$ and $\cL_{b_2}^{(2)}$ are transverse. The number $\mu_{\gamma_1,\gamma_2}(c)$
is the corresponding Maslov index. It is equal to the weighted number of points along $\gamma_2$ between $x_0$ and $c$ where the tangent space to $\cL^{(2)}_{b_2}$ intersect a fiber of $\pi_1$ over a line. The point counts with plus if it is crossed
in the positive direction and with the minus if it is crossed in the negative direction.
When quantization conditions for $b_1$ and $b_2$ hold, the exponent does not depend
on the choice of $\gamma_1$ and $\gamma_2$.
\end{conjecture}

Note that if $M$ is the contangent bundle, this formula is equivalent to the
formula for the scalar product of eigenfunctions of two integrable systems considered
in previous sections.

\subsection{Blattner-Kostant-Sternberg kernels and topological quantum mechanics}
Let $P_1$ and $P_2$ be two real polarizations and $H^{(1/2)}_{P_1}$ and $H^{(1/2)}_{P_2}$
be two corresponding quantization spaces of half-densities with the algebra $C_h(M)$ acting
on them. We can naturally associate a classical integrable system on $M$ with each polarization.
Assume they both have quantizations $C_h(M,B_i)$ where $B_i=M/{P_i}$.

Let $P$ be a third polarization. This polarization gives the representation space
$H^{(1/2)}_{P}$. It is clear that all three spaces should be isomorphic as representations
of $C_h(M)$\footnote{Strictly speaking in such general setting we should conjecture this.}

Scalar products of eigen-half-densities for integrable systems corresponding to $P_1$ and $P_2$
define the unitary linear map  $U_{P_1,P_2}: H^{(1/2)}_{P_1}\to H^{(1/2)}_{P_2}$.
The formula (\ref{general}) describes the semiclassical asymptotic of the integral kernel of
this linear operator. Such integral kernels have
been studied in the context of geometric quantization as well, and are known as
Blattner-Kostant-Steinberg kernels. In terms of half-forms such kernel asymptotically can be written as
\begin{equation}\label{BKS}
U_{P_1,P_2}(b_1,b_2)=\frac{C}{(2\pi h)^{n/2}}\sum_{c\in \cL^{(1)}_{b_1}\cap \cL^{(2)}_{b_2}}
e^{\frac{i}{h}\int_{D_{\gamma_1,\gamma_2}}\omega}\sqrt{\omega^n}(1+O(h))
\end{equation}
The sign of the square is determined by the Malsov index $\mu$ described above.
This formula can be regarded as athe quantization of the symplectomorphisms
$\varphi$ from the remark \ref{genf}. The exponent is exactly the generating
function of the mapping $\varphi$.

The composition law of these integral operators satisfies the semigroup law
and involves the Maslov index. Composing the semiclassical kernels (\ref{BKS})
involves formal integration over the base of intermediate fibration and is a formal
Gaussian computation. The details will be given in a separate publication.

One can argue that this formula correspond to topological quantum mechanics
and can be written as the path integral
\[
U_{P_1,P_2}(b_1,b_2)=\int_{\gamma(0)\in \cL^{(1)}_{b_1}, \gamma(1)\in \cL_{b_2}^{(1)}} e^{\frac{i}{h}\int_\gamma\alpha+f^{(1)}_{b_1}(\gamma(0))-f^{(2)}_{b_2}(\gamma(1))} D\gamma
\]
where $f^{(1)}_{b_1}$ and $f^{(2)}_{b_2}$ are boundary contributions, defined, up to a constant, by the
property $df^{(a)}_{b_a}=\iota_a^*(\alpha)$ where $\iota_a: \cL_{b_a}\hookrightarrow M$ are natural inclusions.
The semiclassical expansion in all orders can be described in terms of
the Poisson sigma model. This is work in progress \cite{CMR}.

\subsection{Geometric asymptotic of $6j$-symbols}

An example of the formula (\ref{general}) describes the Ponzano-Regge  asymptotic of the
Racah-Wigner coefficients, also known as 6j-symbols. For details see \cite{Ro}\cite{TW1}.

Fix a root decomposition for the Lie algebra $su(2)$.
Let $\cO_s\subset su(2)^*$ be the coadjoint orbit for $SU(2)$ passing through the element
$s$ of the dual space ${\mathfrak h}\subset su(2)^*$ to the Cartan subalgebra. Clearly $\cO_s=\cO_{-s}$.
Define the moduli space
\[
\cM_{s_1,s_2,s_3,s_4}=\{(x_1,x_2,x_3,x_4)|x_i\in \cO_{s_i}, x_1+\dots + x_4=0\}/SU(2)
\]
Here the quotient is taken with respect to the diagonal action of $SU(2)$ on the product of
orbits. When $s_1,\dots, s_4$ satisfy triangle inequalities this space is not empty and
$dim( \cM_{s_1,s_2,s_3,s_4})=2$. In this case it is compact, and almost everywhere smooth.
It is clear that the space $\cM_{s_1,s_2,s_3,s_4}$ depends only on the equivalence classes $s_i\to -s_i$
and therefore only on lengths $l_i=|s_i|$ with respect to the metric induced by the Killing form $(-,-)$
on $su(2)^*$.

Level curves of functions
\[
H_{12}=(x_1,x_2), \ \  H_{23}=(x_2,x_3)
\]
define Lagrangian fibrations with fibers
\[
\cL^{(12)}_{l_{12}}=\{(x_1,x_2,x_3,x_4)|x_i\in \cO_{s_i}, x_1+\dots + x_4=0, x_1+x_2\in \cO_{s_{12}}\}/SU(2)
\]
\[
\cL^{(23)}_{l_{23}}=\{(x_1,x_2,x_3,x_4)|x_i\in \cO_{s_i}, x_1+\dots + x_4=0, x_2+x_3\in \cO_{s_{23}}\}/SU(2)
\]
here $l_{12}=|s_{12}|$ is the length of $s_{12}$ and $l_{23}=|s_{23}|$ is the length of $s_{23}$ in the metric defined by the Killing form.
Generic level curves of $H_{12}$ and $H_{23}$ intersect at two points.

The space $V_j$ of an irreducible representation of $su(2)$ can be regarded as the
space of geometric quantization of $\cO_s$. In the semiclassical limit $h\to 0$,  $|s|=\lim (hj)$ as $j\to \infty$.
Similarly, the space of $SU(2)$-invariant vectors $(V_{j_1}\otimes V_{j_2}\otimes V_{j_3}\otimes V_{j_4})^{SU(2)}\subset
V_{j_1}\otimes V_{j_2}\otimes V_{j_3}\otimes V_{j_4}$ can be regarded as the space of geometric quantization of $\cM_{s_1,s_2,s_3,s_4}$ \cite{Ro}\cite{Mar}\cite{TW1}.
In the semiclassical limit $hj_a\to |s_a|$ while $h\to 0$.
Casimir operators acting in $V_{j_1}\otimes V_{j_2}$ and in $V_{j_2}\otimes V_{j_3}$
are quantizations of $H_{12}$ and $H_{23}$ respectively. Denote their eigenhalfdensities
as $\psi_{j_{12}}$ and $\psi_{j_{23}}$ respectively.

The Racah-Wigner coefficients, also known as 6j-symbols, are scalar products $(\psi_{j_{12}},\psi_{j_{23}})$
with respect to the natural scalar product in $V_{j_1}\otimes V_{j_2}\otimes V_{j_3}\otimes V_{j_4}$.
The semiclassical asymptotic of these scalar products was computed using geometric methods
in \cite{Ro} and is an example of (\ref{general}):
\[
(\psi_{j_{12}},\psi_{j_{23}})\simeq \frac{C}{\sqrt{2\pi h}} \left|\frac{\pa^2 S(a,b)}{\pa l_{12}\pa l_{23}}\right|^{1/2}\cos\left(\frac{1}{h}\int_{D_{(a,b)}}\omega +\frac{\pi}{4}\right)(1+O(h))\sqrt{|dl_{12}dl_{23}|}
\]
Here $C$ is an arbitrary constant and $D_{(a,b)}$ is the disc bounded by arcs of $\cL^{(12)}_{l_{12}}$
and $\cL^{(23)}_{l_{23}}$ confined between intersection points $\{a,b\}=\cL^{(12)}_{l_{12}}\cap \cL^{(23)}_{l_{23}}$
For details see \cite{Ro}. All quantities in this formula can be computed explicitly
in terms of the geometry of tetrahedra in 3-dimensional Euclidean space \cite{PR}.  For more details about the semiclassical asymptotic of Racah-Wigner symbols see
\cite{LJ}.

\subsection{Other simple Lie algebras} For simple Lie algebras other then $sl_2$ the decomposition of the tensor product of two irreducible representations typically has multiplicities (for a discussion of 6j-symbols with multiplicities see for example
\cite{Tu}).  But there are special cases of multiplicity free 6j-symbol. One of such examples is the tensor product
of a generic finite dimensional irreducible $sl_n$-module and an irreducible representation with the highest
weight $m\omega_1$ ($m$-th symmetric power of the vector representation of $sl_n$).  The semiclassical limit
of such multiplicity-free 6j-symbols is described by the formula (\ref{general}) and by the geometry of corresponding moduli space.

Let $\g=su_{n}$ be a compact real form of complex Lie algebra $sl_{n}$. Consider four coadjoint orbits $\cO_1,\dots, \cO_4\subset \g^*$ with $\cO_1$ and $\cO_3$ being of rank 1 (orbits corresponding to irreducible $su(n)$ modules with
highest weight $m\omega_1$). The symplectic manifold
 \[
 \cM(\cO_1,\dots, \cO_4)=\{x_i\in \cO_i|x_1+x_2+x_3+x_4=0\}/Ad_G
 \]
 is the symplectic reduction of the product of symplectic manifold
 $\cO_1\times \dots \times \cO_4$ with respect to the diagonal action of $G$.

 It is easy to check that the dimension of $\cM(\cO_1,\dots, \cO_4)$ is $2n-2$. There are two natural
 systems of Poisson commuting Hamiltonians $H_{12}^k=tr((x_1+x_2)^k)$ and $H_{23}^{(k)}=tr((x_2+x_3)^k)$
 where $k=1,\dots, n-1$. Thus, we have two integrable systems. One can show that generic fibers are transversal and intersect at
 $n!$ points.

 The semiclassical asymptotic of 6j symbols in this case is given by the formula  (\ref{general}). This particular multiplicity free case of q-6j symbols is important for the computation of HOMFLY polynomial
 in the semiclassical limit. The combination of the semiclassical limit and the limit $n\to \infty$ was discussed in \cite{AV}.

 \subsection{q-6j symbols}The associativity of the tensor product of representations of $U_q(\mathfrak{g})$ in the
 basis of irreducible components is given by q-6j symbols (see for example \cite{Tu}).
 For $sl_2$ this asymptotic was computed in \cite{TW1} using the difference equation which
 generalizes the computation by Ponzano and Regge.

 One should note that this computation
 is correct only when the signs of the coefficients in this difference equation are suitably stable as
 $h\to 0$.  In this case solutions to the difference equation converge, in the appropriate analytical sense, to solutions of the
 corresponding differential equation.
 For q-6j symbols this means that $q=\exp(\frac{2\pi i}{n})$ and $n\to \infty$. The number of
 irreducible representations of $U_q(sl_2)$  in this case is $n$. This case corresponds to the
 Chern-Simons theory and to the Wess-Zumino unitary conformal field theory.

 When $q$ is another
 root of unity of degree $n$ the analysis based on the difference equation does not work
 and, as it was shown in \cite{CT}\cite{CM}, the asymptotic is not of oscillatory type. This agrees with
 the fact that for such roots of unity the corresponding conformal field theory is not unitary and
 does not correspond to Chern-Simons theory based on a compact simple Lie group.

 When $q=\exp(\frac{2\pi i}{n})$ the asymptotic of q-6j symbols can be expressed in
 terms of the geometry of conjugation orbits for $SU(2)$ and further in terms
 of the geometry of spherical tetrahedra \cite{TW1}.
 For other Lie groups such asymptotic can be naturally computed in terms of cluster variables
 for conjugation orbits \cite{SchSh}. The semiclassical asymptotic of q-6j symbols will be
 analyzed in grater details in a separate paper.

\appendix

\section{Turning points}\label{t-p}

For an integrable system on $T^*Q$ let $\cL_b=H^{-1}(b)$ be the energy surface of a complete set of Hamiltonians. A point $(p_0,q_0)\in \cL_b$ is
{\it simple critical} if the intersection of the tangent plane to $\cL_b$ at this point with the tangent space
to $T^*_{q_0}Q$ is a line. In this case $q_0\in N$ is called a simple turning point.

We assume that simple critical points form a submanifold of dimension $n-1$ in $\cL_b$ which is generically smooth (i.e. that $\cL_b$ is sufficiently generic).

\subsection{One dimensional case} Here we will recall the basic textbook derivation of Maslov indices $\mu$
in the semiclassical formula for eigenhalfdensities.

Let $\widehat{H}=H_h(-ih\frac{\pa}{\pa q}, q)$ be the differential operator with the
principal symbol $H(p,q)$ and $(p_0,q_0)$ be a simple turning point on $\cL_b=\{(p,q)\in T^*Q|H(p,q)=b\}$.
Near this point we have two branches of the level curve $p(q,b)=p_0\pm \sqrt{|\alpha ||q-q_0|}+\dots$
where

$$\alpha=-\frac{\frac{\pa H}{\pa q}(p_0,q_0)}{\frac{\pa^2 H}{\pa p^2}(p_0,q_0)}$$.

\begin{figure}[htb]
\includegraphics[height=5cm,width=8cm]{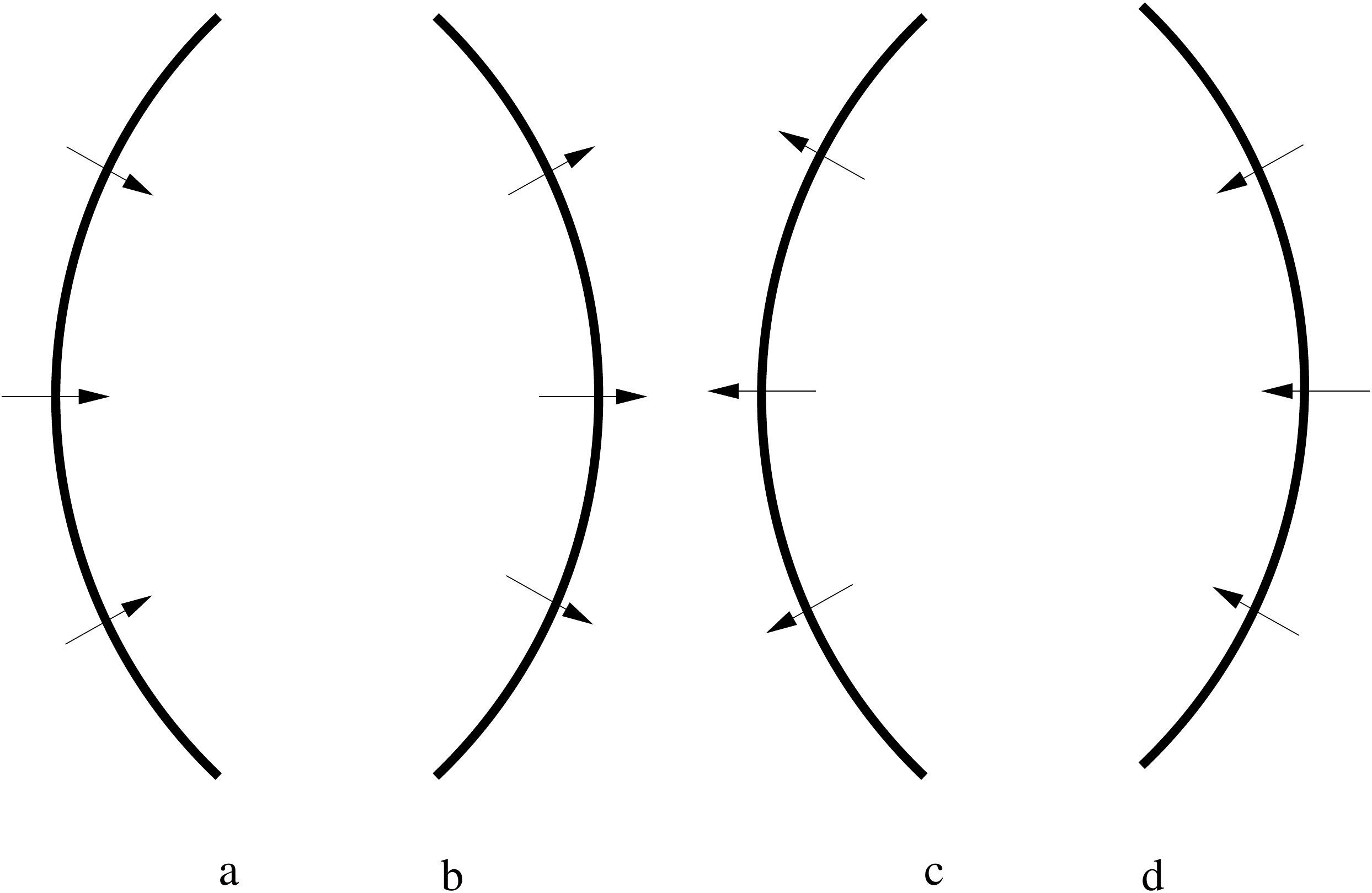}
\caption{Generic turning points $a;b;c;d$ corresponding to $\frac{\pa H}{\pa q}<0, \frac{\pa^2 H}{\pa p^2}>0$;$\frac{\pa H}{\pa q}<0, \frac{\pa^2 H}{\pa p^2}<0$;$\frac{\pa H}{\pa q}>0, \frac{\pa^2 H}{\pa p^2}<0$;$\frac{\pa H}{\pa q}>0, \frac{\pa^2 H}{\pa p^2}>0$ respectively.}
\label{F-3}
\end{figure}

An eigenfunction of $\widehat{H}$ with the eigenvalue $\widehat{b}$ has the following asymptotical behaviour
when $q\to q_0$ and $h\to 0$:
\[
\psi_b(q)=e^{\frac{ip_o(q-q_0)}{h}}\phi(\frac{q-q_0}{h^{2/3}})(1+o(1))
\]
where $\phi(x)$ is an Airy function, i.e. it is a solution to the differential equation
\[
(- \frac{1}{2}\frac{d^2}{dx^2}+\alpha x)\phi(x)=0
\]
Here $\alpha$ is as above. The proof of this fact can be found in various textbooks.


\subsubsection{Airy functions}
These functions are solutions to the differential equation
\[
(-\frac{1}{2} \frac{d^2}{dx^2} +\alpha x)\psi(x)=0
\]
They are given by contour integrals
\[
\psi_C(x)=\int_C e^{i(kx+\frac{k^3}{6\alpha})}dk
\]
Naturally, the integral depends
only on the class of continuous deformations of $C$.

{\bf 1.}$\alpha >0$. The integral is convergent if the
integration contour is approaching to infinity along the rays
\[
k=-it, \ \ k=-it\omega, \ \ k=-it\omega^{-1}
\]
as $t\to +\infty$.

\begin{figure}[htb]
\includegraphics[height=6cm,width=6cm]{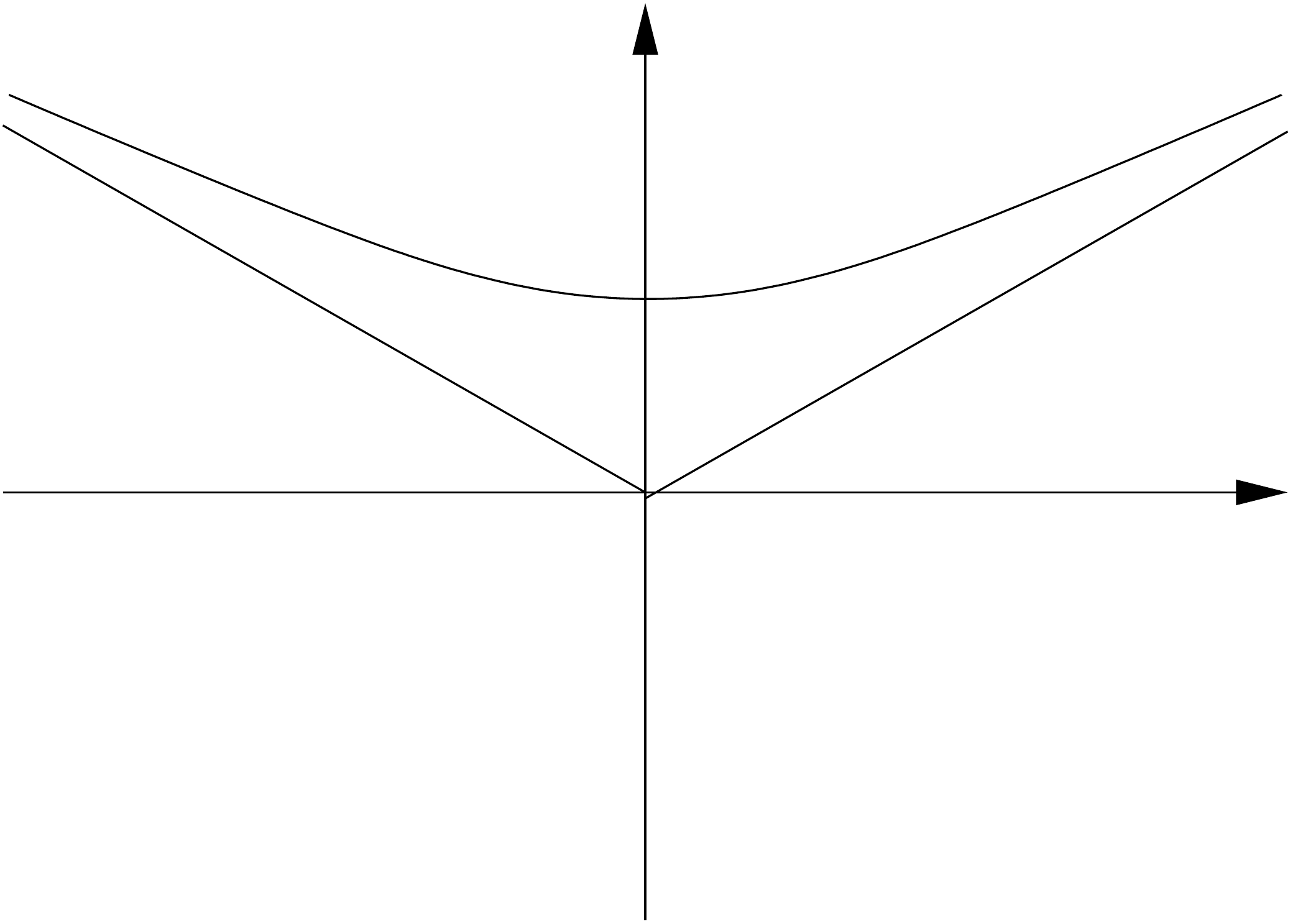}
\caption{The integration contour $C_+$ for $\alpha>0$. The angle between the asymptote and the $x$-axis is $\frac{\pi}{3}$.}
\label{C-1}
\end{figure}

Its asymptotic when $|x|\to \infty$ is determined by critical points
which for $\alpha>0$ are:
\[
k_{1,2}=\pm i\sqrt{\alpha x}, \ \ x\to \infty,
\]
\[
k_{1,2}=\pm\sqrt{\alpha x}, \ \ x\to -\infty,
\]
The solution which has oscillatory asymptotic when $x\to -\infty$ and exponentially decaying
when  $x\to \infty$ correspond to the contour $C_+$ shown in Fig. \ref{C-1}.

The steepest descent method gives the following asymptotic of this function when $x\to \infty$
as the contribution from the critical point $k_1$:
\[
\psi_{C_-}(x)=\sqrt{2\pi } \left( \frac{\alpha}{2x}\right)^{\frac{1}{4}}e^{-\frac{2}{3}\sqrt{2\alpha}x^{\frac{3}{2}}}(1+O(\frac{1}{x}))
\]
The critical point $k_2$ does not contribute.
When $x\to -\infty$ both critical points contribute and the solution $\psi_{C_-}$ has an oscillatory asymptotic:
\[
\psi_{C_-}(x)=\sqrt{2\pi } \left( \frac{\alpha}{2|x|}\right)^{\frac{1}{4}}(e^{-i\frac{2}{3}\sqrt{2\alpha}|x|^{\frac{3}{2}}+i\frac{\pi}{4}}+
e^{i\frac{2}{3}\sqrt{2\alpha}|x|^{\frac{3}{2}}-i\frac{\pi}{4}})(1+O(\frac{1}{|x|}))
\]

\begin{figure}[htb]
\includegraphics[height=6cm,width=6cm]{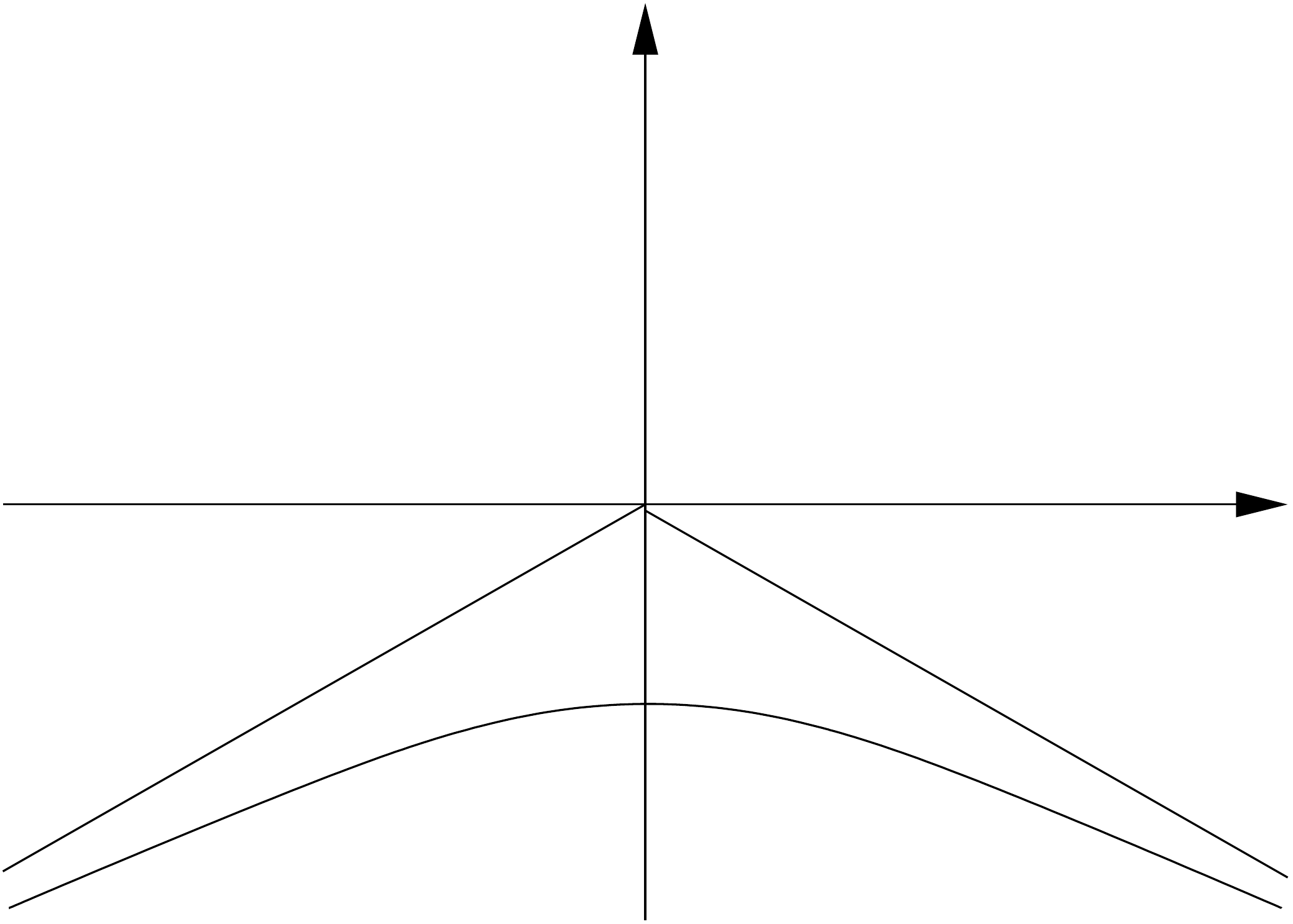}
\caption{The integration contour $C_-$ for $\alpha<0$. The angle between the asymptote and the $x$-axis is $\frac{\pi}{3}$.}
\label{C-2}
\end{figure}

{\bf 2.} Similarly for $\alpha<0$, the contour of integration $C_+$ Fig. \ref{C-2} gives the
solution which exponentially  decays when $x\to -\infty$ and has an oscillatory asymptotic when $x\to \infty$:
\[
\psi_{C_+}(x)=\sqrt{2\pi } \left( \frac{|\alpha|}{2x}\right)^{\frac{1}{4}}(e^{-i\frac{2}{3}\sqrt{2\alpha}x^{\frac{3}{2}}+i\frac{\pi}{4}}+
e^{i\frac{2}{3}\sqrt{2\alpha}x^{\frac{3}{2}}-i\frac{\pi}{4}})(1+O(\frac{1}{x}))
\]
as $x\to \infty$  and
\[
\psi_{C_-}(x)=\sqrt{2\pi } \left( \frac{|\alpha|}{2|x|}\right)^{\frac{1}{4}}e^{-\frac{2}{3}\sqrt{2|\alpha|}|x|^{\frac{3}{2}}}(1+O(\frac{1}{|x|})
\]
as $x\to -\infty$.

\subsubsection{Phase change at turning points} From the asymptotical analysis as $h\to 0$ of solutions to the
eigenvalue problem
\[
\widehat{H}\psi_b(q)=\widehat{b}\psi_b(q)
\]
it is easy to derive that if $\widehat{b}\to b\in B$ we have
\begin{equation}\label{as}
\psi_b(q)=\sum_{a\in \cL_b\cap T^*_qQ} C_{\gamma(a,x_0)} \left|\frac{\pa p(a,b)}{\pa q}\right| e^{\frac{i}{h}S_{\gamma(a,x_0)}(q,b)}(1+O(h))
\end{equation}
Here $x_0$ is a reference point on $\cL_b$. Since $\phi_b(q)$ is defined up to a multiplication by an
arbitrary constant the choice of $x_0$ is not important and only the ratios $C_{\gamma(a,x_0)}/C_{\gamma(c,x_0)}$
are important.

Now let us compute the ratios $C_{\gamma(a,x_0)}/C_{\gamma(c,x_0)}$ by comparing the WKB asymptotic and the Airy asymptotic.

Let $(p_0,q_0)$be a simple turning point. Assume that $q\to q_0$. We have two branches $p_+(q,b)>p_-(q,b)$
with $p_\pm(q_0,b)=p_0$ of the level curve $H(p,q)=b$ over $q$. For small $q$, $p(q,b)=p_0\pm \sqrt{|\alpha ||q-q_0|}+\dots$.
In the neighborhood of this turning point the exponential function in (\ref{wkb-polyn})
corresponding to these branches behaves as
\begin{equation}\label{A2}
S_\pm(q,b)=S(q_0)+p_0(q-q_0)\mp\frac{2}{3}\sqrt{2|\alpha|}|q-q_0|^{3/2}+\dots
\end{equation}
for $\alpha>0$ and $q<q_0$
and
\begin{equation}\label{A1}
S_\pm(q,b)=S(q_0)+p_0(q-q_0)\pm\frac{2}{3}\sqrt{2|\alpha|}(q-q_0)^{3/2}+\dots
\end{equation}
for $\alpha<0$, $q>q_0$.

.

For the prefactors in (\ref{wkb-polyn}) when $q\to q_0$ we have
\[
\left|\frac{\pa p_\pm}{\pa q}\right|=const |q-q_0|^{1/4}(1+o(1))
\]

Let us isolate contributions from the neighborhood of the turning point $q_0$
to the WKB asymptotic:
\[
\psi_b(q)=\left|\frac{\pa p_+(q,b)}{\pa b}\right|^{\frac{1}{4}}e^{\frac{i}{h}S_+(q,b)}C_+\sqrt{|dbdq|}+
\left|\frac{\pa p_-(q,b)}{\pa b}\right|^{\frac{1}{4}}e^{\frac{i}{h}S_-(q,b)}C_-\sqrt{|dbdq|}+\dots
\]
Here "$\dots$" stands for other intersection points of $T^*Q$ and $\cL_b$ and we do not assume that $\psi_b(q)$ is
the WKB asymptotic of eigen-half-density of norm $1$. Taking into account (\ref{A1}) and (\ref{A2})
we obtain the following asymptotic  when  $\alpha>0$ and $q-q_0=h^{2/3}x<0$
\[
\psi_b(q)=A_+\frac{1}{|x|^{1/4}}e^{\frac{i}{h}p_0(q-q_0)}e^{i\frac{2}{3}\sqrt{2\alpha}|x|^{3/2}}+
A_-\frac{1}{|x|^{1/4}}e^{\frac{i}{h}p_0(q-q_0)}e^{-i\frac{2}{3}\sqrt{2\alpha}|x|^{3/2}}+\dots
\]
Here $\dots$ stand for higher order contributions and for the contributions from other
intersection pints in $\cL_b\cap T^*_qQ$. Similarly,
when $\alpha<0$ and $q-q_0=h^{2/3}x>0$ we have
\[
\psi_b(q)=A_+\frac{1}{x^{1/4}}e^{\frac{i}{h}p_0(q-q_0)}e^{-i\frac{2}{3}\sqrt{2|\alpha|}x^{3/2}}+
A_-\frac{1}{x^{1/4}}e^{\frac{i}{h}p_0(q-q_0)}e^{i\frac{2}{3}\sqrt{2|\alpha|}x^{3/2}}+\dots
\]
Here in both cases $A_+/A_-=C_+/C_-$.

Comparing these asymptotics with the asymptotic of Airy functions described before we conclude that
\[
\frac{A_+}{A_-}=\frac{C_+}{C_-}=e^{i\frac{\pi}{2}}
\]
for $\alpha>0$ and
\[
\frac{A_+}{A_-}=\frac{C_+}{C_-}=e^{-i\frac{\pi}{2}}
\]
for $\alpha<0$.

From here we conclude that
\[
\frac{C_{\gamma(a,x_0)}}{C_{\gamma(c,x_0)}}=e^{\frac{i\pi}{2}\mu_{\gamma(a,c)}}
\]
where $\gamma(a,c)$ is the composition of $\gamma(a,x_0)$ and $\gamma(x_0,c)$.
This gives the index $\mu$ in {\ref{wkb-polyn}).

The Bohr-Sommerfeld quantization condition is the consistency condition for (\ref{as}) when coefficients $C_{\gamma(a,x_0)}$
are given by the formula above.

\subsection{Turning points for the cotangent bundle $T^*Q$} In this case the analysis is completely parallel. When $q$ is close to a simple turning point, the Airy analysis in a transversal direction is
completely parallel. We still have $p_+(q,b)$ and $p_-(q,b)$, one is "above" the other \cite{AG}
and the derivation of $\mu$ is literally the same.

\end{document}